\documentstyle[preprint,floats,aps,prl] {revtex}

\begin{document}
\draft

\title{Six-Dimensional Quantum Dynamics of Adsorption and Desorption 
       of H$_2$ at Pd(100):\\ No Need for a Molecular Precursor
       Adsorption State} 

\author{Axel Gross, Steffen Wilke, and Matthias Scheffler}

\address{Fritz-Haber-Institut der Max-Planck-Gesellschaft, Faradayweg 4-6, 
D-14195 Berlin-Dahlem, Germany}

\maketitle

\begin{abstract}

We report six-dimensional quantum dynamical calculations of
dissociative adsorption and associative desorption of the system H$_2$/Pd(100)
using an {\it ab initio} potential energy surface. 
We focus on rotational effects in the steering mechanism, which is
responsible for the initial decrease of the sticking probability with
kinetic energy. In addition, steric effects are briefly discussed.
\end{abstract}


\section{Introduction}

In recent years the dynamics of dissociative adsorption
has been the subject of a large number of experimental and
theoretical investigations (see, e.g., Refs.~\cite{Ren94,Ret95,Hol94,Dar95}).
As far as quantum dynamical simulations were concerned,
these studies were restricted to low-dimensional calculations on model 
potentials due to computational constraints and the non-availability
of {\it ab initio} potential energy surfaces (PES).
By varying potential parameters experimental results were tried to be
reproduced qualitatively. 
These studies laid the foundations
of the current understanding of simple surface reactions and of
the topological features realistic potentials should have. 
The main effects of molecular vibration 
\cite{Jac87,Hal90,Dar92a,Kue91,Sch92,Dar92b}, rotation  
\cite{Mow93,Dar94b,Bru94,Kum94,Dai94,Dai95} and lateral corrugation 
\cite{Dar94,Gro95a} on the dissociative adsorption probability
seemed to be understood to a large extent, but it remained
unclear whether the qualitative explanations would still be
valid in high-dimensional dynamical calculations including all
crucial degrees of freedom.

Just recently it has become possible to evaluate the six-dimensional 
potential energy surface of hydrogen dissociation on metal surfaces
\cite{Ham94,Whi94,Wil95,Wil95b} by density-functional theory.
This development also enforced new efforts for improving the quantum
dynamical algorithms. Indeed it is now feasible to perform studies of hydrogen
dissociation where {\em all} six degrees of freedom of the hydrogen molecule 
are treated quantum mechanically \cite{Gro95b}. 
These calculations showed that the initial decrease
of the sticking probability with kinetic energy found experimentally
for H$_2$ on Pd(100) \cite{Ren89} and on many other transition metal
surfaces \cite{Aln89,Ber92,Res94,Dix94,But94,But95} is not due to
a precursor mechanism, as was commonly believed, but can be explained
by dynamical steering.

In this contribution we will -- after briefly recalling the theoretical
background and the main result of our previous study \cite{Gro95b} --
focus on the influence of rotations on the adsorption dynamics
in the system H$_2$/Pd(100). 
We will describe the dependence of the sticking probability on the initial 
rotational quantum number~$j_i$ of the impinging hydrogen molecules and show 
how this dependence could be verified
experimentally. We end with a brief discussion of steric effects and
concluding remarks.

\section{Theoretical background}

The potential energy surface of H$_2$/Pd(100)
has been determined using the density-functional
theory together with the generalized gradient approximation (GGA) \cite{Per92} 
and the full-potential linear augmented plane wave method \cite{Bla93,Koh95}. 
{\em Ab initio} total energies have been evaluated for more than 250 
configurations and have been parametrized in a suitable form for the 
dynamical calculations \cite{Gro95b}. 
The substrate atoms are assumed to be fixed since due to the large mass 
mismatch between adsorbate and substrate for H$_2$/Pd there is only little 
energy transfer to the substrate phonons. 
The quantum dynamics is determined in a coupled-channel
scheme within the concept of the {\em local reflection matrix} (LORE) 
\cite{Bre93,Chi94} and the {\em inverse local transmission matrix} (INTRA) 
\cite{Bre94}.  This very stable method, which has been employed before 
in a high-dimensional study of the adsorption of H$_2$/Cu(111) \cite{Gro94b},
is closely related to the logarithmic derivative of the solution matrix
and thus avoids exponentially increasing evanescent waves which 
cause numerical instabilities. By utilizing all symmetries of the
hydrogen wave function it has been possible to effectively 
include up to 21,000 channels per total energy in the dynamical calculations.

\section{Results}

\begin{figure}[htb]
\unitlength1cm
\begin{center}
   \begin{picture}(10,6.0)
      \includegraphics{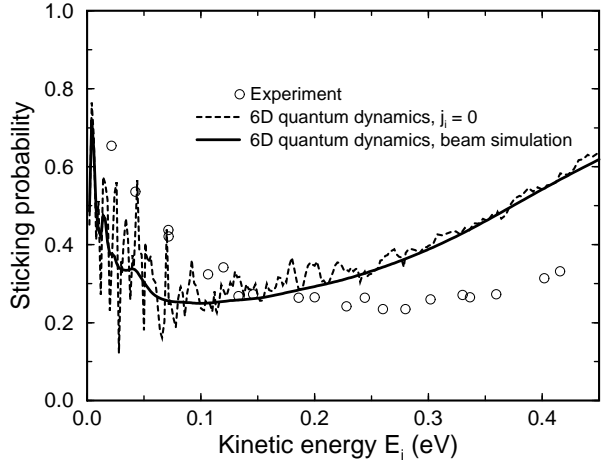}
   \end{picture}

\end{center}
   \caption{Sticking probability versus kinetic energy for
a H$_2$ beam under normal incidence on a Pd(100) surface.
Experiment: circles (from ref.~\protect{\cite{Ren89}}); theory:
H$_2$ molecules initially in the rotational ground state (dashed line)
and with an initial rotational and energy distribution 
adequate for molecular beam experiments (solid line)
(from ref.~\protect{\cite{Gro95b}}). }
\label{stick}
\end{figure}

Figure~\ref{stick} presents the results for the sticking probability as 
a function of the kinetic energy of the H$_2$ beam 
incident on a Pd(100) surface \cite{Gro95b}.
The dashed curve, which corresponds to H$_2$ molecules initially in the 
rotational ground state $j_i = 0$, exhibits a strong oscillatory structure 
for low energies. These oscillations are a consequence of the quantum
nature of the hydrogen beam \cite{Gro95a,Dar90}. They are smoothed out
if the initial rotational population and the energy spread typical for
molecular beam experiments \cite{Ren89} are taken into account (solid line
in fig.~\ref{stick}). This curve should be
compared with the experimental results of Rendulic~{\it et al.} \cite{Ren89}.
The theoretical curve agrees quite well with the experimental data.
Although no precursor state exists in the PES and the energy transfer to 
substrate phonons is not taken into account, 
the initial decrease of the sticking probability with increasing
kinetic energy is well reproduced.

The initial decrease results from a dynamical steering effect
which had been proposed earlier (see, e.g., \cite{Aln89,Dix94}), but
not confirmed theoretically.
Molecules approaching the surface from the gas phase will be attracted
to non-activated paths towards dissociative adsorption by the potential
gradient. The slower the molecules are, the more likely it is that they 
actually follow these attractive paths. By increasing the kinetic energy
the time that the gradient acts upon the molecules is shortened. More 
molecules will then hit the repulsive part of the potential without being 
steered to non-activated paths and will be scattered back into the gas 
phase \cite{Gro95b}. This causes the decrease in the sticking probability.
By further increasing the kinetic 
energy the molecules will eventually have enough energy to directly cross 
the barrier which leads to the increase of the sticking probability at 
higher energies (see fig.~\ref{stick}). In the quantum dynamical
coupled-channel description the steering effect is reflected by the fact that
at low energies more channels are needed in order to get converged results
than at high energies. This indicates that there is 
a strong rearrangement between the different channels at low energies due to
the steering.

\begin{figure}[t]
\unitlength1cm
\begin{center}
   \begin{picture}(10,6)
   \includegraphics{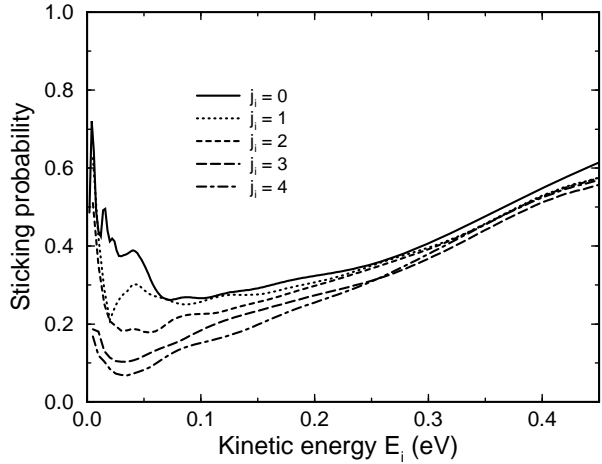}
   \end{picture}
\end{center}
   \caption{Orientationally averaged sticking probability versus kinetic 
            energy for different initial rotational quantum numbers~$j_i$
            of the incoming molecular beam. The molecular beams are assumed
            to have an energy spread of
            $\Delta E / E_i = 2 \Delta v / v_i = 0.2$ \protect{\cite{Ren89}} 
            ($E_i$ and $v_i$ are the initial kinetic energy and velocity, 
            respectively).} 
\label{stickrot}
\end{figure}

The steering occurs in all dynamical degrees of freedom. Therefore by increasing
the energy of, e.g., the rotational degree of freedom of the hydrogen molecule
the steering and thus the sticking probability should be diminished. This can
already be seen in fig.~\ref{stick}, where the sticking probability
of the rotationally populated beam is on the average slightly lower 
as compared to molecules in the rotational ground state. 
This effect is shown in more detail in 
fig.~\ref{stickrot}, which displays the orientationally averaged sticking 
probability  
\begin{equation}\label{ave}
\bar S_{j_i} (E) \ = \ \frac{1}{2j_i +1} \ \sum_{m_i = -j_i}^{j_i} \ 
S_{j_i,m_i} (E), 
\end{equation}
versus initial kinetic energy for $j_i = 0, \ldots , 4$.
Figure~\ref{stickrot} clearly demonstrates that rotational motion hinders
sticking, especially at low kinetic energies, i.e., the regime where
the steering effect is operative.  
The faster the molecules rotate, the more the dissociative adsorption 
is suppressed, because molecules with a high angular momentum will rotate out 
of a favorable orientation towards adsorption during the dissociation event.
This hindering effect of rotations becomes smaller, however, at kinetic 
energies larger than $\sim$0.2~eV, where direct activated adsorption is
dominant.

\begin{figure}[htb]
\unitlength1cm
\begin{center}
   \begin{picture}(10,6)
   \includegraphics{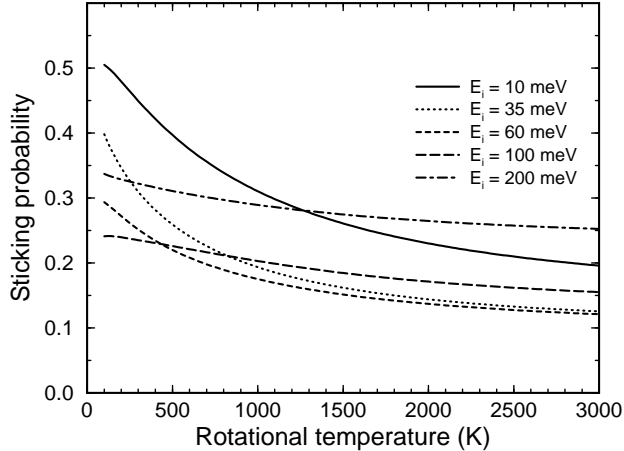}
   \end{picture}
\end{center}
   \caption{Orientationally averaged sticking probability versus rotational
           temperature of the incoming beam for different kinetic energies.} 
\label{rottemp}
\end{figure}

The suppression of the sticking probability by additional rotational 
motion can actually be used to discriminate between the precursor and the
steering mechanism. The precursor state is usually assumed to be a
physisorption state. There are only little directional forces for molecules
adsorbed in a physisorption state, they can almost freely rotate \cite{Zan88}.
The trapping probability into the physisorption state and thus the sticking
probability in the precursor model should be almost
independent of the initial rotational state, in contrast to the steering
mechanism. Unfortunately it is not easy to prepare a molecular beam in
a single quantum state. 
However, by seeding techniques the translational energy of a H$_2$ beam 
can be lowered in a nozzle experiment without changing the rotational 
population of the beam (the translational energy can not be increased since 
there is no lighter seeding gas than H$_2$). In fig.~\ref{rottemp} 
we have plotted the orientationally averaged sticking probability
versus the rotational temperature for different kinetic energies. 
Experimentally the rotational temperature of a H$_2$ beam can not be lower 
than the corresponding translational temperature (a kinetic energy of 200~meV, 
e.g., corresponds to a nozzle temperature of 1200~K), however, theoretically
all combinations of kinetic energy and rotational temperature are feasible. 
For kinetic energies below $\sim$40~meV there is a strong dependence of the
sticking probability on the rotational temperature. By increasing the 
rotational temperature the sticking probability can be decreased by more than
a factor of two at these kinetic energies which should be observable in 
experiment. At large kinetic energies the suppression is
less pronounced which could already be inferred from fig.~\ref{stickrot}.

Interestingly enough, rotational motion seems to suppress sticking 
in general in the system H$_2$/Pd(100). We have checked this for kinetic 
energies $E_i\le$~0.45~eV and rotational quantum numbers $j_i \le$~8.
Also the observed rotational cooling in desorption of H$_2$/Pd(100)
\cite{Gro95b,Sch91} supports these findings. This situation is different in 
the system H$_2$/Cu(111) where a non-monotonous dependence of the sticking 
probability on rotational quantum number $j_i$ has been observed 
\cite{Ret95,Mic93}: Rotational motion is found to hinder adsorption for low 
rotational states ($j_i < 4$) and enhance adsorption for high rotational states
($j_i > 4$) \cite{Ret95}. The enhancement for high $j$ states is related
to the elongation of the molecular bond at the barrier position in the late 
barrier system H$_2$/Cu(111) which leads to a decrease of the rotational 
constant and thus to an effectively lowered barrier for high $j$ states 
\cite{Dar94b,Bru94,Kum94,Dai94,Dai95,Eng93}. 
In the system H$_2$/Pd(100) these late
barriers, however, are absent \cite{Wil95}.

\begin{figure}[htb]
\unitlength1cm
\begin{center}
   \begin{picture}(10,6.0)
      \includegraphics{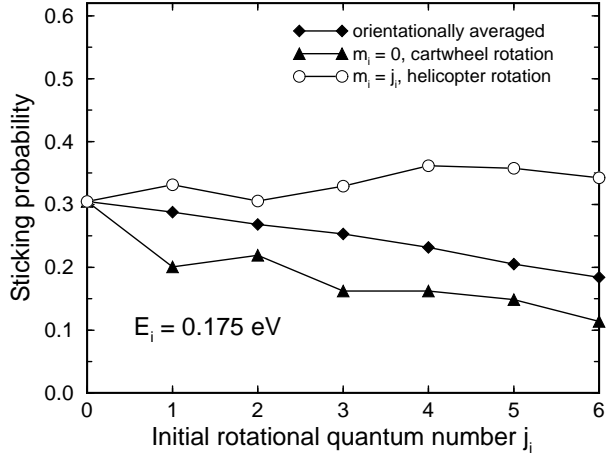}
   \end{picture}

\end{center}
   \caption{Sticking probability versus initial rotational quantum state $j_i$.
   Diamonds: orientationally averaged sticking probability 
   (eq.~\protect{\ref{ave}}),
   triangles: $m_i = 0$ (cartwheel rotation), 
   circles: $m_i = j_i$ (helicopter rotation). 
   The initial kinetic energy is $E_{i} = 0.175$~eV.}

\label{steric}
\end{figure}

There is still an effect that can over-compensate for the suppression
of the sticking probability by rotational motion, namely the orientational 
or steric effect \cite{Gro95b}.  
 The most favorable orientation to adsorption is
with the molecular axis parallel to the surface. Molecules with azimuthal
quantum number~$m = j$ have their axis preferentially oriented parallel to the 
surface. These molecules rotating in the so-called helicopter fashion  
dissociate more easily than molecules rotating in the cartwheel fashion 
($m = 0$) with their rotational axis preferentially parallel to the
surface  since the latter have a high probability hitting the
surface in an upright orientation in which they cannot dissociate. 
This steric effect, which has also been investigated in a number of model 
studies for purely activated adsorption 
\cite{Mow93,Dar94b,Bru94,Kum94,Dai94,Dai95},
can clearly be seen in fig.~\ref{steric} where the sticking probability
for one fixed kinetic energy of $E_i = 0.175$~meV is plotted. Indeed the
$m_i = j_i$~data even rise with increasing quantum number $j_i$
at this relatively high kinetic energy, while the $m_i = 0$ and the
orientationally averaged results are decreasing. At lower kinetic energies
(which is not explicitly shown here), where the steering is more pronounced, 
also the $m_i = j_i$~data decrease.

\vspace{-.5cm}

\section{Conclusions}

In conclusion, we have reported a six-dimensional quantum dynamical
study of adsorption and desorption in the system H$_2$/Pd(100) using
an {\em ab initio} potential energy surface.
We have shown that the initial decrease of the sticking probability 
with increasing kinetic energy is due to dynamical steering. 
We have focused on the steering effect in the rotational
degree of freedom of the hydrogen molecule and shown how the steering
effect can be further confirmed experimentally.
Our study demonstrates that the combination of {\it ab initio}
potential energy surfaces with high-dimensional quantum dynamical
calculations can lead, due to the microscopic information, 
to a {\em quantitative} as well as new {\em qualitative} understanding 
of processes at surfaces.

\vspace{-.5cm}

\section*{Acknowledgements}

We like to thank Prof.~K.~D.~Rendulic for suggesting the detailed
investigation of rotational effects in the steering mechanism.

\end{document}